\documentclass[12pt]{article}

\usepackage[totalwidth=460truept,totalheight=622truept]{geometry}
\usepackage{latexsym,graphicx,amsfonts,amssymb,amsmath,xcolor}
\usepackage[hypertexnames=false,hidelinks]{hyperref}

\def\theequation{\arabic{section}.\arabic{equation}}
\newcommand{\sect}[1]{\setcounter{equation}{0}\section{#1}}
\renewcommand{\theequation}{\thesection.\arabic{equation}}
\global\arraycolsep=1truept
\renewcommand{\theequation}{\arabic{section}.\arabic{equation}}

\newcommand{\be}{\begin{equation}}
\newcommand{\ee}{\end{equation}}

\newcommand{\bs}{\begin{split}}
\newcommand{\es}{\end{split}}

\usepackage[format=hang,font=small,textfont=it]{caption}

\begin{document}

\null

\vskip1truecm

\begin{center}
{\Large \textbf{Classification of $f(R)$ Theories Of Inflation}}

\vskip.6truecm

{\Large \textbf{And The Uniqueness of Starobinsky Model}}

\vskip1truecm

\textsl{{\large Marco Piva}}

\vskip .1truecm

{\textit{Roma, Italy}}

\vspace{0.2cm}

mpiva@fuw.edu.pl
\vskip1truecm

\vskip .5truecm

\textbf{Abstract}
\end{center}

We classify $f(R)$ theories using a mathematical analogy between slow-roll inflation and the renormalization-group flow. We derive the power spectra and spectral indices class by class and compare them with the latest data. The framework used for the classification allows us to determine the general structure of the $f(R)$ functions that belong to each class. Our main result is that only two classes survive.
Moreover, we show that the Starobinsky model is the only polynomial $f(R)$ that can realize slow-roll inflation. In fact, all other polynomials belong to a special class that can only realize constant-roll inflation, at least far enough in the past. We point out some of the issues involved in considering a smooth transition between constant-roll and slow-roll inflation in this class of models. Finally, we derive the map that transforms the results from the Jordan frame to the Einstein frame.

\vfill\eject

\section{Introduction}
Inflation is one of the best paradigms used to solve the main puzzles of primordial cosmology: the horizon problem and the flatness problem. In particular, it strikes for its simplicity. Indeed, it is sufficient to extend Einstein gravity with the $R^2$ term to explain the present data and solve the above problems, as shown by Starobinsky~\cite{Starobinsky:1980te}. Alternatively, the same result can be achieved by introducing an additional scalar field with a suitable potential~\cite{Guth:1980zm, Linde:1981mu}, without modifying general relativity.

During the last decades these two possibilities have been largely studied and extended by considering different  potentials~\cite{Martin:2013tda} or by generalizing the Starobinsky model to the so-called $f(R)$ theories~\cite{Sotiriou:2008rp, DeFelice:2010aj}, where a general function of the scalar curvature $R$ is used in the gravitational action. Moreover, the same models have been further extended by using the Palatini formulation of gravity that, for a certain class of theories, gives different physical predictions compared to the standard metric formulation (see e.g.~\cite{Bauer:2008zj} and~\cite{Olmo:2011uz} for a review). In the end, the simplicity of the solution offered by inflation has been traded for a plethora of models that are all compatible with the data and suitable to solve the open problems. Therefore, until we do not have other strong data to falsify most of these models, it is compulsory to include some fundamental principles that can help us navigate and select theories out. A good lesson can be learned from particle physics, where power counting, renormalizability, unitarity and gauge symmetries strongly constrain the  models. In this regard, it is remarkable that Starobinsky inflation is obtained by including the only term with dimension 4 that can be added to the Hilbert action without spoiling unitarity. This points in the direction that power counting, renormalizability and unitarity might play important roles in primordial cosmology. It is also possible to make the Starobinsky model renormalizable by adding the gravitational action all terms with dimension 4~\cite{Stelle:1976gc}. This includes the cosmological term and the $R_{\mu\nu}R^{\mu\nu}$ term. However, the latter causes a violation of unitarity due to the presence of a massive spin-2 ghost. Such unwanted degree of freedom can be removed from the spectrum without renouncing to renormalizability by turning it into a so-called “purely virtual" particle~\cite{Anselmi:2018ibi, Anselmi:2018tmf}, thus restoring unitarity. However, in the context of cosmology the presence of the $R_{\mu\nu}R^{\mu\nu}$ term affects only the perturbations~\cite{Anselmi:2020lpp}, while the background dynamics stays unmodified. Nevertheless, a sharp prediction for the tensor-to-scalar ratio has been derived~\cite{Anselmi:2020lpp}.

Recently, a new way to classify single-field inflationary models has been proposed~\cite{Anselmi:2021rye}. It is based on a mathematical analogy between the renormalization-group (RG) flow in quantum field theory and the evolution of the background metric during inflation~\cite{Anselmi:2020shx}, where the role of the coupling constant is played by an appropriate slow-roll parameter $\alpha$.  In this framework, a key quantity is the analogous of the beta function $\beta_{\alpha}$, which encodes the background dynamics in a perturbative way. In particular,  it is required to be a power series in $\alpha$ that starts from a quadratic term with negative coefficient. Then, the various scalar-field potentials can be classified based on their beta functions. However, in  in~\cite{Anselmi:2021rye} the study of $f(R)$ was left out. The reason is that, moving to the Einstein frame, each $f(R)$ generates a different and rather complicated potential and, in some cases, it is not even possible to determine them analytically. Moreover, since the definition of $\alpha$ is arbitrary, in the Einstein frame it is nontrivial to find the right one to study the $f(R)$ altogether. The only exception is the potential associated to the Starobinsky model, which was indeed included in the analysis of~\cite{Anselmi:2021rye}.

In this paper we extend the classification of~\cite{Anselmi:2021rye} to $f(R)$ theories directly in the Jordan frame, which allows us to identify the correct definition for $\alpha$. In particular, assuming that the beta function be a power series, we classify the models and derive the leading predictions class by class without specifying the form of $f(R)$. Moreover, starting from a beta function with generic coefficients, we work out the expression of all the $f(R)$ that belong to the same class in terms of such coefficients.

Comparing the results with the most up-to-date experimental constraints we find conditions on the first nonvanishing coefficient of the beta functions and exclude all classes but two. Remarkably, if we focus on polynomial $f(R)$, which in general relativity are generated by renormalization anyway, we find that the only one suitable for slow-roll inflation is the Starobinsky model. In fact, extending the analysis to theories with a linear beta function, it turns out that all the other polynomials belong to such a class. A particular property of these theories is that they cannot realize slow-roll inflation but only \emph{constant-roll inflation}~\cite{Motohashi:2014ppa}, i.e. an accelerated expansion with a constant rate. We highlight some difficulties in interpolating between slow-roll and constant-roll inflation, and, thus, between the Starobinsky model and its polynomial extensions. 

Finally, once we derive all the quantities in the Jordan frame we provide the map to the Einstein frame for comparison with~\cite{Anselmi:2021rye}.

The paper is organized as follows. In~\autoref{sect:fRG} we review $f(R)$ theories, introduce the RG analogy and find the definition of $\alpha$ that simplifies the equations. In~\autoref{sec:pert} we derive the quadratic action for tensor and scalar perturbations in a generic $f(R)$ theory and recall the derivation of the spectra used in~\cite{Anselmi:2021rye}. In~\autoref{sec:predictions} we work out the leading-order predictions for the spectral indices and the tensor-to-scalar ratio and show that every class that does not have quadratic or cubic beta function is ruled out by data. In~\autoref{sec:classes} we derive the general form of the $f(R)$ that belongs to the surviving classes and their power spectra. In~\autoref{sec:frame} we derive the perturbative map between the Jordan frame and the Einstein frame. In~\autoref{sec:class0} we study a special class of theories with linear beta function and point out the issues with these models. Finally,~\autoref{sec:conclusions} contains our conclusions.

\emph{Notation and conventions:} We use the signature $(+,-,-,-)$ for the metric tensor. The Riemann and Ricci tensors are defined as $R^{\mu}_{ \ \nu\rho\sigma}=\partial_{\rho}\Gamma^{\mu}_{\nu\sigma}-\partial_{\sigma}\Gamma^{\mu}_{\nu\rho}+\Gamma^{\mu}_{\alpha\rho}\Gamma^{\alpha}_{\nu\sigma}-\Gamma^{\mu}_{\alpha\sigma}\Gamma^{\alpha}_{\nu\rho}$ and $R_{\mu\nu}=R^{\rho}_{ \ \mu\rho\nu}$, respectively. In the case of cosmological perturbations we omit the integral over the space momentum and use the word “action" for the integral over time of the lagrangian density.

\sect{$f(R)$ theories and the cosmic RG flow}
\label{sect:fRG}
In this section we briefly review  $f(R)$ theories, introduce the quantities used to define the cosmic RG flow directly in the Jordan frame and explain our strategy for classifying the theories. 

We start by recalling the action and equations of motion on the Friedmann-Lema\^itre-Robertson-Walker (FLRW) background. The starting action reads
\be\label{eq:fRact}
S(g)=-\frac{1}{2\kappa^2}\int\sqrt{-g}f(R),\qquad \kappa^2=8\pi G,
\ee
where $G$ is the Newton constant. The equations of motion are
\be
F(R)R_{\mu\nu}-\frac{1}{2}g_{\mu\nu}f(R)+(g_{\mu\nu}\nabla^2-\nabla_{\mu}\nabla_{\nu})F(R)=0, \qquad F(R)\equiv\frac{\partial f}{\partial R}.
\ee
Choosing the FLRW background
\be
g_{\mu\nu}^{\text{FLRW}}\mathrm{d}x^{\mu}\mathrm{d}x^{\nu}=\mathrm{d}t^2-a(t)^2\mathrm{d}x^{i}\mathrm{d}x^{i}
\ee
we get
\be
H \dot{F}-F\dot{H}-FH^2-\frac{f}{6}=0,
\ee
\be
\ddot{F}+2\dot{H}F-H\dot{F}=0,
\ee
where $H\equiv\dot{a}/a$ is the Hubble constant. 
Now we adapt the procedure of~\cite{Anselmi:2021rye} to $f(R)$ theories. The ingredients to define a RG flow are as follows:
\begin{itemize}
\item[a)] a small quantity (analogous of the coupling constant in quantum field theory);
\item[b)] a first-order differential equation for that small quantity (analogous of the beta function);
\item[c)] correlation functions that are conserved along the flow (analogous of the Callan-Symanzik equation).
\end{itemize}

Since during inflation the background metric must be approximately de Sitter, an obvious small quantity would be the slow-roll parameter
\be
\varepsilon\equiv -\frac{\dot{H}}{H^2}.
\ee
Indeed, this is a good choice if we work with a specific $f(R)$ model (see~\cite{Anselmi:2020lpp}). However, we would like to treat these types of theories on general grounds. It turns out that another useful quantity is instead  
\be\label{eq:coupling}
\alpha\equiv-\frac{\dot{F}}{2HF},
\ee
which we show below to be small during inflation. Following~\cite{Anselmi:2021rye} we also define
\be\label{eq:vbeta}
v\equiv-\frac{1}{Ha\tau}, \qquad \beta_{\alpha}\equiv\frac{\mathrm{d}\alpha}{\mathrm{d}\ln|\tau|},
\ee
where $\tau$ is the conformal time. The beta function $\beta_{\alpha}$ is a key quantity to classify the theories and gives us a first order differential equation for $\alpha$ to define the cosmic RG flow. In fact $\beta_{\alpha}$ encodes the evolution of the background metric during the inflationary phase in a way that resembles that of coupling constants in quantum field theories and their behavior at different energies. It is also useful to turn every time derivative into $\alpha$-derivative by means of the relation
\be\label{eq:dt}
\frac{\mathrm{d}}{\mathrm{d}t}=-vH\beta_{\alpha}\frac{\mathrm{d}}{\mathrm{d}\alpha}.
\ee

In terms of~\eqref{eq:coupling} and~\eqref{eq:vbeta} the equations of motion become
\be\label{eq:betaeom}
-v\beta_{\alpha}=\alpha(1+2\alpha)-(1-\alpha^2)\varepsilon,
\ee
\be\label{eq:varepsilon}
\varepsilon=1+2\alpha+\frac{f}{6FH^2}.
\ee
In this framework, we want to derive all the background quantities as expansions in $\alpha$ assuming that $\beta_{\alpha}$ be a power series. First, we discuss its general form and find the properties that are necessary to successfully realize slow-roll inflation. To do so, we introduce another common parameter
\be
\eta_H\equiv\frac{\dot{\varepsilon}}{\varepsilon H}.
\ee
For slow-roll inflation we usually require
\be\label{eq:SRconditions}
\varepsilon,|\eta_H|\ll 1.
\ee
Therefore, we need to translate the conditions~\eqref{eq:SRconditions}
into properties of $\beta_{\alpha}$. Using~\eqref{eq:betaeom}, the parameters $\varepsilon$ and $\eta_H$ are
\be\label{eq:epseta}
\varepsilon=\frac{\alpha+2\alpha^2+v\beta_{\alpha}}{1-\alpha}, \qquad \eta_H=-\frac{v\beta_{\alpha}\left[2+4\alpha+v\left(\frac{\mathrm{d}\beta_{\alpha}}{\mathrm{d}\alpha}-1\right)\right]}{\alpha+2\alpha^2+v\beta_{\alpha}}.
\ee
Since in  exact de Sitter spacetime $v=1$, we need a beta function that is at least quadratic in order to have both $\varepsilon$ and $\eta_H$ of at least order $\mathcal{O}(\alpha)$, i.e.
\be
\beta_{\alpha}=\alpha^2\sum_{n=0}^{\infty}b_n\alpha^n,
\ee
where the $b_i$ are constant coefficients. Note that if we admit a linear beta function, then $\varepsilon=\mathcal{O}(\alpha)$, while $\eta_H=\mathcal{O}(\alpha^0)$. In this case slow-roll inflation cannot be realized because $\eta_H$ is constant in the far past. This class of models is discussed in~\autoref{sec:class0}.

As mentioned above, we could have chosen $\varepsilon$ as a small quantity to define the RG flow. In that case we would have had
\be
\eta_H=-\frac{v\beta_{\varepsilon}}{\varepsilon}.
\ee
Again, assuming $\varepsilon\ll 1$, we need that $\beta_{\varepsilon}$ be at least quadratic to guarantee $\eta_{H}=\mathcal{O}(\varepsilon)$. Since $\varepsilon(\alpha)$ is a simple power series, it means that also $\alpha(\varepsilon)$ is. We conclude that assuming $\alpha\ll 1$ and $\beta_{\alpha}=\mathcal{O}(\alpha^2)$ is equivalent to assume $\varepsilon\ll 1$ and $\beta_{\varepsilon}=\mathcal{O}(\varepsilon^2)$. This means that, during the inflationary phase, the roles of $\alpha$ and $\varepsilon$ are interchangeable without loss of generality. Passing from one to the other is just a perturbative redefinition. 

Finally, the background spacetime needs to reach the de Sitter one in the asymptotic past. This means that the first nonvanishing coefficient of the beta function must be negative. Because of these properties, the whole construction resembles that of beta functions and running-coupling constants in quantum field theory, where an asymptotically-free theory requires a negative beta function. Here the ``flow" is represented by the background evolution during the perturbative phase of inflation. More details can be found in~\cite{Anselmi:2021rye}. It is important to highlight that such an analogy is just a mathematical correspondence and has nothing to do with quantum corrections, which are not considered in this paper.

Now we derive differential equations for each quantity and solve them perturbatively in $\alpha$. Depending on the class, the solutions might factorize overall nonpolynomial functions. From the definition of $v$ we find
\be\label{eq:veps}
\beta_{\alpha}\frac{\partial v}{\partial\alpha}=1-v-\varepsilon.
\ee
Equation~\eqref{eq:veps} can be solved by a simple power series 
\be
v(\alpha)=\sum_{n=0}^{\infty}v_n\alpha^n.
\ee
Then from the definitions of $\alpha$ and $\varepsilon$ we get $F(\alpha)$ and $H(\alpha)$ in terms of $b_i$ by solving the equations
\be\label{eq:FHeqs}
v\beta_{\alpha}\frac{\partial F}{\partial\alpha}=2\alpha F, \qquad v\beta_{\alpha}\frac{\partial H}{\partial\alpha}= H\varepsilon.
\ee
Finally, once we have $v$, $F$, $H$ and $\varepsilon$ we can use~\eqref{eq:varepsilon} to determine $f$ as a function $\hat{f}$ of $\alpha$ as
\be\label{eq:feq}
\hat{f}(\alpha)=6FH^2\left(\varepsilon-1-2\alpha\right).
\ee
Then we perturbatively invert $R(\alpha)$ from 
\be\label{eq:RHeq}
R(\alpha)=-6H^2(2-\varepsilon)
\ee
to find $\alpha(R)$ and obtain
\be
f(R)=\hat{f}(\alpha(R)).
\ee
We say that two $f(R)$ theories belong to the same ``class" if their $\beta_{\alpha}$ have the same leading order. Therefore, if we follow the procedure explained above assuming generic beta functions, then we  obtain the $f(R)$ that belong to the same class.

We call \emph{class I} the set of theories with quadratic beta function, \emph{class II} with cubic one and so on. In~\autoref{sec:classes} we make this derivation for class I and class II, showing which type of $f(R)$ belong to each class.

As explained above, the definition of the beta function gives a first-order differential equation for $\alpha$. In a fashion similar to that of quantum field theory, we introduce a scale $k$ with mass dimension 1 through the definition of the \emph{rescaled conformal time} $\eta$
\be
\eta\equiv-k\tau,
\ee
which, in our RG analogy, play the role of the renormalization scale, together with its logarithm $\ell$
\be
\ell\equiv \ln\eta.
\ee
Then the equation for $\alpha$ reads 
\be\label{eq:eqalpha}
\frac{\mathrm{d}\alpha(\ell)}{\mathrm{d}\ell}=\beta_{\alpha}(\alpha).
\ee
Its solution give a sort of \emph{running slow-roll parameter}.
It is useful to solve the equation for $\alpha$ with the initial condition $\alpha(\ell=0)=\alpha_k$, which means at $\tau=-1/k$. The idea is to later identify $k$ with the modulus of the space momentum of perturbations. In this way, once we know $\alpha_k$ at a certain fixed scale $k_*$, we can evolve it to any other scale $k$. Indeed, as in the standard renormalization group, the fact that the scale $k$ is introduced by hand allows us to write the equation
\be\label{eq:alphakeq}
\frac{\mathrm{d}\alpha}{\mathrm{d}\ln k}=0, \qquad \Rightarrow \qquad \frac{\mathrm{d}\alpha_k}{\mathrm{d}\ln k}=-\beta_{\alpha}(\alpha_k).
\ee
Equation~\eqref{eq:eqalpha} can be solved by truncating the beta function. Since its formal expressions do not depend on the frame, we can use the same listed in~\cite{Anselmi:2021rye}. For class I and II at the leading-log order (LLO) we have
\be\label{eq:alpharun12}
\alpha^{\text{I}}(\ell)=\frac{\alpha_k}{1-b_0\alpha_k\ell}, \qquad \alpha^{\text{II}}(\ell)=\frac{\alpha_k}{\sqrt{1-2b_1\alpha_k^2\ell}}.
\ee
We do not report the next-to-leading-log (NLL) corrections since in this paper we can directly use the results of~\cite{Anselmi:2021rye}. However, it is important to highlight that all the results for power spectra and spectral indices will eventually be written as expansions in $\alpha_k$, which is fixed at a pivot scale $k_*$ by the experimental data. Then thanks to the second equation in~\eqref{eq:alphakeq} we can evolve $\alpha$ to an arbitrary $k$. For example, in class I we have 
\be
\alpha_k=\frac{\alpha_{k_*}}{1+b_0\alpha_{k_*}\ln\left(k/k_*\right)}.
\ee
Therefore, our results include the leading-log corrections that are already resummed in the running slow-roll parameter and are usually computed with much more effort with the standard approaches.

Finally, the last ingredient for the RG analogy is an equation of the Callan-Symanzik type for correlators. This is given by the conservation of power spectra $\mathcal{P}$ in the superhorizon limit $\eta\rightarrow 0$
\be
\frac{\mathrm{d}\mathcal{P}}{\mathrm{d}\ln|\tau|}=0.
\ee
Since $\mathcal{P}$ can be viewed as a function of $\alpha(\ell)$ and $\alpha_k$, its conservation on superhorizon scales means that the dependence on $\alpha$ drops and $k$ appears only through $\alpha_k$. More detail can be found in~\cite{Anselmi:2020shx}.

\sect{Perturbations}
\label{sec:pert}

In this section we outline the general strategy to derive the quadratic action for the perturbations and solve their equations of motion by briefly reviewing the method used in~\cite{Anselmi:2021rye}. However, we can directly export the solutions of the Mukhanov-Sasaki equation as well as the expressions for the power spectra and spectra indices at the NNLL and NLL for tensor and scalar perturbations, respectively. The reason is that in~\cite{Anselmi:2021rye} those results were derived for a generic action quadratic in the perturbations, without specifying its origin. 

Cosmological perturbations are studied by introducing small deviations from the FLRW metric, which are then decomposed according to the irreducible representations of 3-dimensional rotations. We work with (spatial) Fourier-transformed quantities and assume that the space momentum $\textbf{k}$ is oriented along the $z$ axis. Then we perform a change of field variables and switch to the rescaled conformal time $\eta= -k\tau$, where $k=|\mathbf{k}|$, so the final action quadratic in the redefined perturbation $w$ is in the Mukhanov-Sasaki form, i.e.
\be\label{eq:MSact}
S_2(w)=\frac{1}{2}\int\mathrm{d}\eta\left[w'^2-w^2+\frac{2+\sigma}{\eta^2}w^2\right],
\ee
where $\sigma$ is a function that depends on background quantities. Typically~\cite{Anselmi:2021rye}, $\sigma$ is analytic in $b_i$, which means that it can only depend on $v$ and $\beta_{\alpha}$, as well as on $\alpha$ explicitly, because $H$, $F$, $f$ and $R$ are different for each class and not analytic in the first nonvanishing $b_i$.

The Mukhanov-Sasaki equation for the perturbation $w$ reads

\be\label{eq:MSeq}
w''+w-\frac{2+\sigma}{\eta^2}w=0.
\ee
In the framework adopted here, we expand $w(\eta)$ in powers of $\alpha_k$ and solve~\eqref{eq:MSeq} perturbatively, i.e.
\be\label{eq:wexpansion}
w(\eta)=\sum_{n=0}^{\infty}w_{n}(\eta)\alpha_k^n.
\ee
The functions $w_{n}(\eta)$ are determined by inserting~\eqref{eq:wexpansion} in~\eqref{eq:MSeq}, which gives the set of equations
\be\label{eq:eqwn}
w''_{n}+w_{n}-\frac{2}{\eta^2}w_{n}=\frac{g_{n}(\eta)}{\eta^2},
\ee
where $g_n$ are functions that depend on $\sigma$ and on the $w_m$ with $m<n$. The Bunch-Davies condition
\be\label{eq:bd}
w(\eta)\rightarrow\frac{e^{i\eta}}{\sqrt{2}}, \qquad \text{for} \qquad \eta\rightarrow\infty.
\ee
is imposed by requiring that
\be
w_0(\eta)\rightarrow \frac{e^{i\eta}}{\sqrt{2}}, \qquad  w_{n>0}(\eta)\rightarrow 0 \qquad \text{for} \qquad \eta\rightarrow\infty.
\ee
Finally, the solution in terms of $f_n$ is~\cite{Anselmi:2020shx}
\be
w_{n}(\eta)=\int_{\eta}^{\infty}\mathrm{d}\eta'\frac{g_{n}(\eta')}{\eta\eta^{\prime 3}}\left[(\eta-\eta')\cos (\eta-\eta')-(1+\eta\eta')\sin (\eta-\eta')\right].
\ee

The procedure just explained gives the $w$ as an explicit function of $\eta$. However, it is possible to single out the part of $w$ (more precisely of $\eta w$) that survives the superhorizon limit. Such part depends on $\ln\eta$ only. For this reason, it can be written as a function of $\alpha$. We summarize the steps here below. More details can be found in~\cite{Anselmi:2020shx} and ~\cite{Anselmi:2021rye}. 

First we decompose $w$ as 
\be\label{eq:wdec}
\eta w(\eta)=Q(\eta)+Y(\eta),
\ee
where $Q$ is a function of $\ln\eta$, while $Y$ goes to zero in the superhorizon limit, i.e. for $\eta\rightarrow 0$.  Plugging back~\eqref{eq:wdec} in~\eqref{eq:MSeq} and using the properties of $Q$ and $Y$, we can find an equation for $Q$, which reads
\be\label{eq:Qeq}
\frac{\mathrm{d}Q}{\mathrm{d}\ln\eta}=-\frac{1}{3}\frac{1}{1-\frac{1}{3}\frac{\mathrm{d}}{\mathrm{d}\ln\eta}}
\sigma Q=-\frac{\sigma Q}{3}-\frac{1}{3}\sum_{n=1}^{\infty}3^{-n}\frac{\mathrm{d}^n(\sigma Q)}{\mathrm{d}\ln^n\eta}.
\ee
Then, $Q$ can be viewed as a function $\tilde{Q}(\alpha,\alpha_k)$ of $\alpha$ and $\alpha_k$
\be
\tilde{Q}(\alpha,\alpha_k)=Q_0(\alpha_k)\frac{J(\alpha)}{J(\alpha_k)}
\ee
 and equation~\eqref{eq:Qeq} turns into
 \be\label{eq:Jeq}
  \frac{\mathrm{d}J}{\mathrm{d}\alpha}=-\frac{\sigma J}{3}-\frac{1}{3}\sum_{n=1}^{\infty}3^{-n}\left(\beta_{\alpha}\frac{\mathrm{d}}{\mathrm{d}\alpha}\right)^n(\sigma J).
 \ee
From~\eqref{eq:Jeq} we can derive $J$ as a power series in $\alpha$ times an overall function, depending on the class. Finally, from the limit $\eta\rightarrow 0$ we obtain $Q_0$ 
\be
Q(0)= \lim_{\eta\rightarrow 0} \eta w(\eta)=Q_0(\alpha_k).
\ee
In this way we can identify a perturbation (up to an arbitrary constant $C$)
\be\label{eq:wc}
w^{\text{C}}(\eta)\equiv\frac{C}{k^{3/2}}\frac{\eta w(\eta)}{J(\alpha)},
\ee
that is conserved in the superhorizon limit. Indeed,
\be
\lim_{\eta\rightarrow 0}w^{\text{C}}=\frac{C}{k^{3/2}}\frac{Q_0(\alpha_k)}{J(\alpha_k)}.
\ee
Both in the tensor and scalar case, the constant $C$ can be chosen such that $w^{\text{C}}$ coincides with the usual gauge-invariant perturbations, which are also conserved in the superhorizon limit. The power of this method is that it gives a recipe to obtain gauge-invariant perturbations that are conserved on superhorizon scales by construction. Although this result might not seems surprising in the case of single-field inflation, it becomes particularly relevant in the case of multifield inflation, where it is nontrivial to find conserved perturbations. See~\cite{Anselmi:2021dag} and~\cite{Grzadkowski:2025yxr} for applications of the RG techniques in the context of double-field inflation.

After defining the quantum perturbation as the operator
\be
\hat{w}^{\text{C}}_{\mathbf{k}}(\eta)=w_{\mathbf{k}}^{\text{C}}(\eta)\hat{a}_{\mathbf{k}}+w_{\mathbf{-k}}^{\text{C}*}(\eta)\hat{a}^{\dagger}_{\mathbf{-k}}
\ee
with 
\be
\left[\hat{a}_{\mathbf{k}},\hat{a}^{\dagger}_{\mathbf{k}'}\right]=(2\pi)^3\delta^{(3)}(\mathbf{k}+\mathbf{k}'),
\ee
the power spectrum $\mathcal{P}_{w^{\text{C}}}$ is obtained from
\begin{equation}
\langle \hat{w}_{\mathbf{k}}^{\text{C}}(\eta )\hat{w}_{\mathbf{k}^{\prime
}}^{\text{C}}(\eta )\rangle =(2\pi )^{3}\delta ^{(3)}(\mathbf{k}+\mathbf{k}%
^{\prime })\frac{2\pi ^{2}}{k^{3}}\mathcal{P}_{w^{\text{C}}},
\end{equation}
while its spectral index is 
\be
n_{w}-\theta=\frac{\mathrm{d}\ln\mathcal{P}_{w^{\text{C}}}}{\mathrm{d}\ln k}=-\beta_{\alpha}(\alpha_k)\frac{\partial \ln\mathcal{P}_{w^{\text{C}}}}{\partial \alpha_k},
\ee
where $\theta=0,1$ for tensor and scalar perturbations, respectively. In the superhorizon limit we have
\be\label{eq:PSsuperhorizon}
\mathcal{P}_{w^{\text{C}}}\simeq\frac{C^2}{2\pi^2}\frac{\left|Q_0(\alpha_k)\right|^2}{J(\alpha_k)^2}.
\ee
Finally, the tensor-to-scalar ratio is
\be
r(k)\equiv \frac{\mathcal{P_{\text{t}}}(k)}{\mathcal{P}_{\text{t}}(k)},
\ee
where $\mathcal{P_{\text{t}}}$ and $\mathcal{P_{\text{s}}}$ are the tensor and scalar power spectra, respectively.

As anticipated in the beginning of this section, since in~\cite{Anselmi:2021rye} the functions $w_n$, as well as $J$ and $Q_0$, have been computed assuming a power series for $\sigma$ with generic coefficients,  we can directly use those results. The expressions for $w_n$ are analytic in $b_i$, while those for $J$, and therefore those for power spectra, depend on the class. The reason is that the transformation from $w$ back to the original perturbation involves $F$, $H$ and $f$. 
 
In what follows, we work out the quadratic action for each type of perturbation and derive the corresponding expression for $\sigma$. We use the subscripts “t" and “s" to distinguish the various quantities in the two cases.

\subsection{Tensors}
We start from tensor perturbations, which are studied  by writing

\be
g_{\mu\nu}=g_{\mu\nu}^{\text{FLRW}}+\delta g_{\mu\nu},\qquad
\delta g_{\mu\nu}=-2a^2\left(\setlength\arraycolsep{5pt}\begin{array}{cccc}
0 & 0 & 0 & 0\\
0 & u & \tilde{u} & 0\\
0 & \tilde{u} & -u & 0\\
0 & 0 & 0 & 0\\
\end{array}\right),
\ee
where $u=u(t,z)$ and $\tilde{u}=\tilde{u}(t,z)$ are the graviton modes.
After performing the spatial Fourier transform, the action quadratic in $u$ reads
\be
S_u=\frac{1}{\kappa^2}\int\mathrm{d}t a^3F\left(\dot{u}^2-\frac{k^2}{a^2}u^2\right).
\ee
The perturbation $\tilde{u}$ gives an identical action. We consider only the action for $u$ and accounts for the contribution of $\tilde{u}$, as well as for the polarizations, by defining the tensor power spectrum as
\be
\mathcal{P}_{\text{t}}\equiv 16 \mathcal{P}_{u},
\ee
where $\mathcal{P}_{u}$ is the power spectrum of $u$.
From now on, it is understood that $u=u_{\mathbf{k}}(t)$ and $u^2=u_{\mathbf{k}}u_{-\mathbf{k}}$, $\dot{u}^2=\dot{u}_{\mathbf{k}}\dot{u}_{-\mathbf{k}}$, where $u_{\mathbf{k}}(t)$ is the Fourier transform of $u(t,z)$.

In order to reduce the action in the form~\eqref{eq:MSact} we apply the redefinition 
\be\label{eq:tensred}
u=\frac{\kappa w_{\text{t}}}{a\sqrt{2kF}}
\ee
and switch to the rescaled conformal time. Then, after some manipulations, the new action is in the form~\eqref{eq:MSact} with
\be\label{eq:sigmat}
\sigma_{\text{t}}=\frac{2(1-v^2)-\alpha(4+\alpha)}{v^2},
\ee
which depends only on $v$ as promised.

\subsection{Scalars}
In order to simplify the study of scalar perturbations, we write a new action by introducing an auxiliary field $\phi$ in the following way
\be\label{eq:actphi}
S'(g,\phi)=-\frac{1}{2\kappa^2}\int\sqrt{-g}\left[f(\phi)-f'(\phi)\left(\phi-R\right)\right].
\ee
The equation of motion for $\phi$ gives
\be
f''(\phi)\left(\phi-R\right)=0.
\ee
It is straightforward to see that plugging the solution $\phi(R)=R$ into~\eqref{eq:actphi} gives back the original action~\eqref{eq:fRact}, i.e. 
\be
S'(g,\phi(R))=S(g).
\ee
Note that $\phi$ is just an auxiliary field and its introduction is not a change of frame.

We choose to work in the spatially-flat gauge, where the scalar perturbations of the metric are parametrized by
\be
\delta g_{\mu\nu}=\left(\setlength\arraycolsep{5pt}\begin{array}{cccc}
2\Phi(t,z) & 0 & 0 & -a\partial_zB(t,z)\\
0 & 0 & 0 & 0\\
0 & 0 & 0 & 0\\
-a\partial_zB(t,z) & 0 & 0 & 0\\
\end{array}\right).
\ee
Moreover, we write
\be
\phi=\bar{R}+\delta\phi(t,z), \qquad \bar{R}=-6H^2(2-\varepsilon),
\ee
where $\delta\phi$ is a small perturbation of $\phi$. With this parametrization the quadratic action is
\be
S_2(\Phi,B,\delta\phi)=\frac{1}{\kappa^2}\int\mathrm{d}ta^3\mathcal{L}_2
\ee
with
\be
\begin{split}
\mathcal{L}_2=&\frac{\delta \phi ^2 F_1}{4}+3 H \Phi  \left[\Phi  F H (2 \alpha -1)+\dot{\delta\phi} F_1\right]+3 H \delta \phi 
   \Phi  \left[\dot{F_1}+H (\varepsilon -1) F_1\right]\\
&+\frac{k^2}{a^2}\left\{\delta \phi  \Phi  F_1-a B \left[2 F H\Phi (\alpha -1) +\dot{\delta\phi} F_1+\delta \phi  \left(\dot{F_1}-H
   F_1\right)\right]\right\},
   \end{split}
   \ee
where $F_1\equiv\frac{\partial F}{\partial R}$. We can see that the fields $\Phi$ and $B$ are auxiliary ones. In particular, the equation of motion for $B$ gives $\Phi$
\be
\frac{\delta S_2}{\delta B}=0\qquad\Rightarrow\qquad  \Phi=\frac{\dot{\delta\phi}F_1+\delta\phi\left(\dot{F}_1-HF_1\right)}{2 FH(1-\alpha)},
\ee
which once plugged back into the action removes $B$, leaving the action
\be
S'_2(\delta\phi)=\frac{3}{4\kappa^2}\int\mathrm{d}t\frac{a^3F_1^2}{F(\alpha-1)^2}\left(\dot{\delta\phi}^2-\Omega\delta\phi^2\right),
\ee
where $\Omega$ is a complicated function of background quantities that we do not report. After performing the change of variable
\be\label{eq:scalred}
\delta\phi=\kappa\sqrt{\frac{2F}{3 k}}\frac{(\alpha-1)}{a F_1}w_{\text{s}}
\ee
and some manipulations, we switch to the rescaled conformal time $\eta$ and obtain an action in the form~\eqref{eq:MSact} with
\be\label{eq:sigmas}
\sigma_{\text{s}}=\sigma_{\text{t}}-\frac{\beta_{\alpha}}{ \alpha}\left[\frac{2}{v}+\frac{1}{1-\alpha}\left(1-\frac{\mathrm{d}\beta_{\alpha}}{\mathrm{d}\alpha}-\frac{2\beta_{\alpha}}{1-\alpha}\right)\right].
\ee
Also in this case, $\sigma_{\text{s}}$ depends only on $v$ and $\beta_{\alpha}$.
From the expressions of $\sigma_{\text{t,s}}$ we can derive their expansion in $\alpha$ and apply the results of~\cite{Anselmi:2021rye}.

\sect{Predictions}
\label{sec:predictions}
In this section derive the leading-order predictions for the spectral indices and the tensor-to-scalar ratio assuming a generic class. In the next section we derive all the expansions in detail for explicit classes. 

First, we recall that the power spectra in the superhorizon limit read
\be\label{eq:PtPssh}
\mathcal{P}_{\text{t}}\simeq\frac{8C_{\text{t}}^2}{\pi^2}\frac{\left|Q_{0\text{t}}(\alpha_k)\right|^2}{J_{\text{t}}(\alpha_k)^2}, \qquad \mathcal{P}_{\text{s}}\simeq\frac{C_{\text{s}}^2}{\pi^2}\frac{\left|Q_{0\text{s}}(\alpha_k)\right|^2}{J_{\text{s}}(\alpha_k)^2}.
\ee
The arbitrary coefficients $C_{\text{t,s}}$ can be chosen such that $w^{\text{C}}_{\text{t}}(\eta(t))=u(t)$ and $w^{\text{C}}_{\text{s}}(\eta(t))=\mathcal{R}\equiv H\frac{\delta\phi}{\dot{\bar{R}}}$, where $\mathcal{R}$ is the curvature perturbation in the spatially-flat gauge. Indeed, the quantities $w^{\text{C}}_{\text{t}}$ and $w^{\text{C}}_{\text{s}}$ can only be proportional to $u$ and $\mathcal{R}$, respectively, being gauge invariant and conserved on superhorizon scales. Comparing the redefinitions~\eqref{eq:tensred} and~\eqref{eq:scalred} with the expression~\eqref{eq:wc} we find
\be\label{eq:Jts}
J_{\text{t}}=\frac{C_{\text{t}}\sqrt{2 F}}{\kappa v H}, \qquad J_{\text{s}}=\frac{C_{\text{s}}\alpha\sqrt{6 F}}{\kappa v H(1-\alpha)},
\ee
hence
\be\label{eq:generalPtPs}
 \mathcal{P}_{\text{t}}=\frac{4\kappa^2v(\alpha_k)^2H(\alpha_k)^2\left|Q_{0\text{t}}(\alpha_k)\right|^2}{\pi^2F(\alpha_k)}, \qquad \mathcal{P}_{\text{s}}=\frac{\kappa^2v(\alpha_k)^2H(\alpha_k)^2(1-\alpha_k)^2\left|Q_{0\text{s}}(\alpha_k)\right|^2}{12\pi^2\alpha_k^2F(\alpha_k).}.
\ee

Then, using equations~\eqref{eq:veps} and~\eqref{eq:FHeqs}, the tensor-to-scalar ratio and the spectral indices read
\be\label{eq:generalr}
r=48\left(\frac{\alpha_k}{1-\alpha_k}\right)^2\left|\frac{Q_{0\text{t}}(\alpha_k)}{Q_{0\text{s}}(\alpha_k)}\right|^2,
\ee

\be
n_{\text{t}}=2\left[1-\frac{1-\alpha_k}{v(\alpha_k)}-\beta_{\alpha}(\alpha_k)\frac{\mathrm{d}\ln|Q_{0\text{t}}(\alpha_k)|}{\mathrm{d}\alpha_k}\right],
\ee

\be
n_{\text{s}}-1=2\left\{1-\frac{1-\alpha_k}{v(\alpha_k)}+\beta_{\alpha}(\alpha_k)\left[\frac{1}{\alpha_k(1-\alpha_k)}-\frac{\mathrm{d}\ln|Q_{0\text{s}}(\alpha_k)|}{\mathrm{d}\alpha_k}\right]\right\}.
\ee
It is interesting to note that, since $r$ and $n_{\text{t,s}}$ depend only on $v$, $\beta_{\alpha}$ and $Q_{0\text{t,s}}$, they are analytic\footnote{From~\cite{Anselmi:2021rye} we know that also $Q_{0\text{t,s}}$ are analytic in $b_i$.} in $b_i$. 

With the expressions found above we can derive the leading-order contributions assuming a generic beta function 
\be
\beta_{\alpha}=b_n\alpha^{2+n}+\mathcal{O}(\alpha^{3+n}), \qquad n\in \mathbb{N}.
\ee
Using that the solution of~\eqref{eq:veps} is
\be
v(\alpha)=1-\alpha-3\alpha^2+\mathcal{O}(\alpha^3), 
\ee
and that
\be 
Q_{0\text{t}}(\alpha_k)=\frac{i}{\sqrt{2}}+\mathcal{O}(\alpha_k^2),\qquad Q_{0\text{s}}(\alpha_k)=\frac{i}{\sqrt{2}}+\mathcal{O}(\alpha_k).
\ee
we find
\be
r=48\alpha_k^2+\mathcal{O}(\alpha_k^3)
, \qquad \forall n\geq 0,
\ee
\be
n_{\text{t}}=-6\alpha_k^2+\mathcal{O}(\alpha_k^3), \qquad \forall n\geq 0,
\ee

\be\label{eq:nsLO}
n_{\text{s}}-1=\begin{cases}
2\alpha_kb_0+\mathcal{O}(\alpha_k^2) & \text{if } n=0,\\
2\alpha_k^2(b_1-3)+\mathcal{O}(\alpha_k^3)  & \text{if } n=1,\\
-6\alpha_k^2 +\mathcal{O}(\alpha_k^3) & \text{if } n \geq 2.
\end{cases}
\ee
These results are formally the same found in~\cite{Anselmi:2021rye} for single-field inflation in the Einstein frame.

We can now compare the results with data. The most up-to-date experimental constraints are obtained by combining the results from the Planck 2018 collaboration~\cite{Planck:2018jri}, the dark energy spectroscopic instrument~\cite{DESI:2024uvr} and recent data from the Atacama Cosmology Telescope~\cite{ACT:2025fju,ACT:2025tim}. The result gives
\be\label{eq:nsexp}
n_{\text{s}}(k_*)=0.974\pm 0.003
\ee
for $k_*=0.05 \ \text{Mpc}^{-1}$. From now on every quantity with the subscript ``$*$" is understood to be at $k=k_*$. From~\eqref{eq:nsexp} we can derive $\alpha_*$ for each class and plug it back in the leading order of the tensor-to-scalar ratio $r_*$. In this way we obtain a relation between $r_*$ and $n_{\text{s}*}$. We find
\be\label{eq:rstar}
r_*\simeq\begin{cases}
\frac{12(1-n_{\text{s}*})^2}{b_0^2} & \text{if } n=0,\\
\frac{24(1-n_{\text{s}*})}{3-b_1} & \text{if } n=1,\\
8(1-n_{\text{s}*})  & \text{if } n \geq 2.
\end{cases}
\ee
Using the central value for $n_{\text{s}*}$ and taking into account the bound for the tensor-to-scalar ratio $r_*<0.035$ from the Bicep/KECK collaboration~\cite{BICEP:2021xfz}, we find that the accepted values for the first nonvanishing coefficient of the beta functions are
\be\label{eq:constrbn}
b_0\lesssim-0.48, \qquad b_1\lesssim-14.8,
\ee
while any class higher than II is ruled out since it would give $r_*\simeq 0.208$, which is too large. Therefore, we can consider only $n=0,1$.

In the literature, the number of e-folds before the end of inflation $N_k$ is often used as a primary constraint, since there are various theoretical estimates that place it between 50 and 60. However, those derivations depend on the inflationary model, as well as on the reheating temperature (see~\cite{Gorbunov:2011zzc} for a clear derivation of the estimate and~\cite{Zharov:2025evb} for a recent study). Moreover, the number of e-folds is also a frame-dependent quantity. For these reasons, we prefer to fix the free parameters of the models by means of the experimental data for $n_{\text{s}}$ and derive the prediction for $N_k$, instead of imposing $N_k$ in a range and then check the prediction for $n_{\text{s}}$. Then the reader can make their own interpretations of the results using the values for $N_k$ that they consider acceptable.

The number of e-folds before the end of inflation is defined as

\be\label{eq:Nk}
N_k=\int_{t_k}^{t_f}\mathrm{d}tH(t)=-\int_{\alpha_k}^{\alpha_f}\frac{\mathrm{d}\alpha'}{v(\alpha')\beta_{\alpha}(\alpha')},
\ee
where $t_f$ is the time when inflation ended, while $t_k$ is the time when perturbations with momentum $k$ exited the horizon. At the leading order it reads

\be
N_k\simeq-\int_{\alpha_k}^{\alpha_f}\frac{\mathrm{d}\alpha'}{b_n\alpha'^{n+2}}=-\frac{1}{b_n(n+1)\alpha_k^{1+n}}+\mathcal{O}(\alpha_k^0).
\ee
Using~\eqref{eq:nsLO}
for class I and II we get

\be
N_*\simeq\begin{cases}
\frac{2}{1-n_{\text{s}*}} & \text{if } n=0,\\
\frac{1}{1-n_{\text{s}*}}\left(1-\frac{3}{b_1}\right) & \text{if } n=1.
\end{cases}
\ee
Using again the central value of $n_{\text{s}*}$ and the constraints~\eqref{eq:constrbn} we have

\be
N_*^{n=0}\simeq 77\pm 9,\qquad 
N_*^{n=1}\lesssim 47\pm 7.
\ee
 If we accept that $N_*$ can be slightly larger than 60, then any $f(R)$ in class I is in agreement with data, provided that $b_0\lesssim-0.48$. On the other hand, if we impose $50\leq N_*\leq 60$ then class I is almost out of the two-sigma range. However, this is a mild tension and it is not enough to exclude the models in class I, considering that there are also some tensions between the data themselves~\cite{DESI:2025gwf,Zharov:2025evb}. 
 Functions of class II are in good agreement at the one-sigma range.

Finally, we highlight that $f(R)$ theories can only be considered as effective theories and therefore have a limited validity. Potential tension could be resolved within more fundamental theories, such as renormalizable ones.

\sect{Derivation of $f(R)$ from $\beta_{\alpha}$}
\label{sec:classes}
In this section we follow the strategy explained in~\autoref{sect:fRG} to obtain the expansions of all the background quantities for class I and II. In this way we can identify the general expression for the $f(R)$ that belong to those classes. Finally, we use the results of~\cite{Anselmi:2021rye} to work out the power spectra.

First we need $v, F$ and $H$. In principle, their expansions could factorize an overall, nonpolynomial function of $\alpha$. To check if it is the case, we solve the differential equations~\eqref{eq:veps} and~\eqref{eq:FHeqs} at the leading order by assuming different betas.

As anticipated in~\autoref{sect:fRG}, the equation for $v$ can be solved by a power series with coefficients that are analytic in $b_i$. Therefore, its expression is valid for any class and reads

\be
\begin{split}\label{eq:vexpansion}
v(\alpha)=1-&\alpha -3 \alpha ^2-\alpha ^3 \left(3-6 b_0\right)-3 \alpha ^4 \left(1-4 b_0+6 b_0^2-2 b_1\right)\\
&-3 \alpha
   ^5 \left[1+18 b_0^2-24 b_0^3-b_0 \left(6-14 b_1\right)-4 b_1-2 b_2\right]+\mathcal{O}(\alpha^6).
   \end{split}
\ee
Using the expression for $v$ we can write the expansions for~\eqref{eq:sigmat} and~\eqref{eq:sigmas}, which are also analytic in $b_i$,

\be\label{eq:sigmatexp}
\begin{split}
\sigma_{\text{t}}=&9 \alpha ^2+6 \alpha ^3 \left(3-4 b_0\right)+3 \alpha ^4 \left(21-24 b_0+24 b_0^2-8 b_1\right)\\
&+12 \alpha ^5
   \left[15+24 b_0^2-24 b_0^3-b_0 \left(27-14 b_1\right)-6 b_1-2 b_2\right]+\mathcal{O}(\alpha^6),
   \end{split}
\ee

\be\label{eq:sigmasexp}
\begin{split}
\sigma_{\text{s}}=-3 \alpha  b_0+\alpha ^2 \left(9-3 b_0+2 b_0^2-3 b_1\right)+\alpha ^3 \left[4 b_0^2+b_0 \left(5 b_1-33\right)-3
   \left(b_1+b_2-6\right)\right]+\mathcal{O}(\alpha^4).
   \end{split}
\ee

From the coefficients of $\sigma_{\text{t,s}}$ we can directly obtain the solutions of the Mukhanov-Sasaki equation in the two cases using~\cite{Anselmi:2021rye}. The results up to the order $\mathcal{O}(\alpha_k^3)$ for tensors and $\mathcal{O}(\alpha_k^2)$ for scalars are
\be
w_{\text{t}0}=W_0, \qquad w_{\text{t}1}=0, \qquad w_{\text{t}2}=W_2, \qquad w_{\text{t}3}=2W_2-\frac{b_0}{2}W_3.
\ee

\be
w_{\text{s}0}=W_0, \qquad w_{\text{s}1}=-\frac{b_0}{3}W_2, \qquad w_{\text{s}2}=\frac{1}{3}\left(b_0+b_1-3+\frac{7}{6}b_0^2\right)W_2+\frac{b_0^2}{4}W_4,
\ee
with
\begin{eqnarray}
W_{0} &=&\frac{i(1-i\eta )}{\eta \sqrt{2}}\mathrm{e}^{i\eta },\qquad W_{2}=3%
\left[ \text{Ei}(2i\eta )-i\pi \right] W_{0}^{\ast }+\frac{6W_{0}}{(1-i\eta )%
},  \notag \\
W_{3} &=&\left[ 6(\ln \eta +\tilde{\gamma}_{M})^{2}+24i\eta
F_{2,2,2}^{1,1,1}\left( 2i\eta \right) +\pi ^{2}\right] W_{0}^{\ast }+\frac{%
24W_{0}}{(1-i\eta )}-4(\ln \eta +1)W_{2},  \label{Wi} \\
W_{4} &=&-\frac{16W_{0}}{1+\eta ^{2}}+\frac{2(13+i\eta )W_{2}}{9(1+i\eta )}+%
\frac{W_{3}}{3}+4G_{2,3}^{3,1}\left( -2i\eta \left\vert _{0,0,0}^{\
0,1}\right. \right) W_{0},  \notag
\end{eqnarray}%
where Ei denotes the exponential-integral function, $F_{b_{1},\cdots
,b_{q}}^{a_{1},\cdots ,a_{p}}(z)$ is the generalized hypergeometric function 
$_{p}F_{q}(\{a_{1},\cdots ,a_{p}\},\{b_{1},\cdots ,b_{q}\};z)$ and $%
G_{p,q}^{m,n}$ is the Meijer-G function, while $\tilde{\gamma}_M=\gamma_M-\frac{i\pi}{2}=\gamma_{E}+\ln 2-\frac{i\pi}{2}$, $\gamma_E$ being the Euler-Mascaroni constant. 

Now we derive the expansions that are class dependent.
\subsection{Class I}

Choosing $b_0\neq0$ the equations~\eqref{eq:FHeqs} cannot be solved by simple power series. We find that the solutions are
\be
\frac{F(\alpha)}{\alpha^{2/b_0}F_0}=1+2 \alpha  \left(\frac{1}{b_0}-\frac{b_1}{b_0^2}\right)+\alpha ^2 \left[\frac{4}{b_0}+\frac{\left(b_1-4\right) b_1}{b_0^3}+\frac{2 b_1^2}{b_0^4}+\frac{2-b_1-b_2}{b_0^2}\right]+\mathcal{O}(\alpha^3),
\ee

\be
\frac{H(\alpha)}{\alpha^{1/b_0}H_0}=1+\alpha  \left(1+\frac{4}{b_0}-\frac{b_1}{b_0^2}\right)+\alpha ^2
   \left[1+\frac{9}{b_0}+\frac{\left(b_1-8\right) b_1}{2 b_0^3}+\frac{b_1^2}{2 b_0^4}+\frac{16-6 b_1-b_2}{2
   b_0^2}\right]+\mathcal{O}(\alpha^3),
\ee
where $F_0$ and $H_0$ are constants with $[F_0]=0$ and $[H_0]=1$.
Using~\eqref{eq:RHeq} we can write the expansion for $R$ and then invert it. We find 
\be
\begin{split}
\alpha(R)=X\Bigg[1-&X \left(4+\frac{3 b_0}{4}-\frac{b_1}{b_0}\right)\\
&+X^2 \left(19+\frac{149 b_0}{16}+\frac{35 b_0^2}{32}-\frac{9
   b_1}{4}+\frac{b_1^2}{b_0^2}+\frac{b_2-20 b_1}{2 b_0}\right)+\mathcal{O}(X^3)\Bigg]
   \end{split}
\ee
where we have defined the quantity $X$ as
\be
X=\tilde{R}^{-n}, \qquad n\equiv -2b_0,\qquad \tilde{R}\equiv-\frac{R}{12H_0^2}.
\ee
Since at the leading order the scalar curvature is
\be
R(\alpha)\simeq-12H_0^2\alpha^{2/b_0}
\ee
and $b_0<0$, we have that around de Sitter spacetime $X$ is positive and small. The expression for $f(R)$ is
\be\label{eq:fclass1}
f(R)=-6F_0H_0^2\tilde{R}^2\left[1+\frac{3}{n \tilde{R}^n}+\frac{18-36 n+13 n^2-3 b_1}{4 n^2 \tilde{R}^{2 n}}+\mathcal{O}(\tilde{R}^{-3n})\right].
\ee
We see that it is possible to obtain $f(R)$ with inverse and/or fractional powers of $\tilde{R}$. However, those models can only be phenomenological ones, since they do not have a well-defined Minkowski limit and/or they could not be used on every spacetime because $\tilde{R}$ could be negative in general. 

In order to have more usable theories, we impose some constraints on the function. First, there should be no singularity for $\tilde{R}=0$, in order to have a well-defined Minkowski limit, as well as any other singularity for real $\tilde{R}$. Then we also avoid terms involving $\sqrt{\tilde{R}}$. These considerations force $n$ to be a positive integer\footnote{A priori we could consider $n\in\mathbb{Q}$ and then fix the relations between the $b_i$ such that fractional powers of $R$ do not appear. However, this is not possible since the term proportional to $\tilde{R}^{-n}$ in the square bracket of~\eqref{eq:fclass1} cannot be set to zero.}. Such constraints are useful to obtain functions that are suitable in every spacetime background. However, if we are interested in functions that are guaranteed to be well defined around de Sitter spacetime only, we can relax those assumptions (see the end of this subsection for an example).

The simplest choices are polynomials. In this class there are only two possibilities. Here we write them in a more familiar way after fixing the $b_i$ and  redefining $F_0$ and $H_0$. The details are reported in~\autoref{app:examples}. We find
\be
f(R)=R-\frac{1}{6M^2}R^2+2\Lambda, \qquad f(R)=\frac{1}{6M^2}R^2+2M^2.
\ee
The first is the Starobinsky model, while the second is pure quadratic gravity. In both cases a cosmological constant is present. However, in the latter it cannot be set to zero. Moreover, the $R^2$ term has the wrong sign and the theory propagates a ghost. This can be fixed by choosing a negative $F_0$, which would then lead to a negative cosmological constant. Therefore, the second model is not physically acceptable and we are left only with the Starobinsky one.

Nonpolynomial functions can be obtained by keeping infinitely many terms and choose $b_i$ such that they reproduce the Taylor-Laurent series of some known function. For example, we can have
\be
f(R)=-\frac{R^2}{6 M^2}\left(1+\frac{3}{2}\frac{M^4}{M^4+R^2}\right),
\ee
which nonsingular for real $R$.

Finally, if we relax one of the constraints on $f(R)$ and allow for terms containing powers of $\sqrt{\tilde{R}}$ we can have modifications to the Starobinsky model, such as 
\be
f(R)=R-\frac{1}{6M^2}R^2-\frac{\lambda}{M}(-R)^{3/2},\qquad \lambda>0,
\ee
where $[\lambda]=0$. This model was studied in~\cite{Ketov:2010qz}.  

Using formulas from~\cite{Anselmi:2021rye}, we obtain the following expressions for power spectra and spectral indices

\be
\mathcal{P}^{\text{I}}_{\text{t}}=\frac{4C_{\text{t}}^2}{\pi^2}\left[1+\frac{6 \alpha _k}{b_0}+6 \alpha _k^2 \left(1-\gamma_M-\frac{1}{ b_0}+\frac{6-b_1}{2b_0^2}\right)+\mathcal{O}(\alpha_k^3) \right],
\ee

\be
n_{\text{t}}^{\text{I}}=-6 \alpha _k^2 \left[1-2\alpha_k\Big(1+b_0(1-\gamma_M)\Big)\right]+\mathcal{O}(\alpha^4_k),
\ee

\be
\begin{split}
\mathcal{P}^{\text{I}}_{\text{s}}=\frac{C_{\text{s}}^2}{4\pi^2\alpha^2_k}\Bigg\{1-2 \alpha _k \left[1-\frac{3}{b_0}+(2-\gamma_M) b_0\right]+\mathcal{O}(\alpha_k^2)\Bigg\},   \end{split}
   \ee
 
\be
n_{\text{s}}^{\text{I}}-1=2 \alpha _k \left\{b_0+\alpha _k \left[b_0+(2-\gamma_M) b_0^2+b_1-3\right]+\mathcal{O}(\alpha_k^2)\right\},
\ee

\be
r^{\text{I}}=48\alpha_k^2\left\{1+2 \alpha_k  \left[1+b_0 \left(2-\gamma _M\right)\right]+\mathcal{O}(\alpha_k^2)\right\}.
\ee

The coefficient $C_{\text{t,s}}$ can be obtained from~\eqref{eq:Jts} and using the explicit expressions for $F$, $H$ and $J_{\text{t,s}}$. We find
\be
C_{\text{t}}=\frac{\kappa H_0}{\sqrt{2 F_0}}, \qquad C_{\text{s}}=\frac{\kappa H_0}{\sqrt{6 F_0}}=\frac{C_{\text{t}}}{\sqrt{3}}.
\ee
This result is the same for every class. Note that the relation between $C_{\text{s}}$ and $C_{\text{t}}$ is the same as in single-field inflation with any potential~\cite{Anselmi:2021rye}.

\subsection{Class II}
In general, the higher the class the more complicated are the singularities in the expressions of $F$ and $H$ at $\alpha=0$. Indeed, choosing $b_0=0$ and $b_1\neq 0$, the solutions of equations~\eqref{eq:FHeqs} are
\be
F(\alpha)=e^{-\frac{2}{\alpha  b_1}} \alpha ^{\frac{2 \left(b_1-b_2\right)}{b_1^2}} F_0 \left[1+2\alpha\left(\frac{4}{b_1}+\frac{b_2^2}{b_1^3}-\frac{b_2+b_3}{b_1^2}\right)+\mathcal{O}(\alpha^2)\right],
\ee

\be
H(\alpha)=e^{-\frac{1}{\alpha  b_1}} \alpha ^{\frac{4 b_1-b_2}{b_1^2}} H_0 \left[1+\alpha 
   \left(1+\frac{10}{b_1}+\frac{b_2^2}{b_1^3}-\frac{4 b_2+b_3}{b_1^2}\right)+\mathcal{O}(\alpha^2)\right],
\ee
which have essential singularities. In this case inverting $R(\alpha)$ gives an expansion in powers of $1/\mathcal{W}_0$, where $\mathcal{W}_0$ is the first branch of the Lambert function, defined as 
\be
\mathcal{W}_0(x)e^{\mathcal{W}_0(x)}=x,
\ee
for some variable $x$. Indeed, we find
\be
\alpha(R)=X\left[1-X^2 \left(10+\frac{3 b_1}{4}+\frac{b_2^2}{b_1^2}-\frac{4 b_2+b_3}{b_1}\right)+\mathcal{O}(X^3)\right],
\ee
where
\be
X=\frac{\gamma}{b_1}\frac{1}{\mathcal{W}_0\left(\frac{\gamma}{b_1}\tilde{R}^{-\gamma/2}\right)}, \qquad \tilde{R}=-\frac{R}{12H_0^2},\qquad \gamma\equiv \frac{b_1^2}{4b_1-b_2},
\ee
which produces rather complicated $f(R)$'s. However, we can simplify the result by choosing some particular subclasses. For example, if we set $b_2=4b_1$ and redefine for convenience $b_1=-6n$ with $n>0$, the quantity $X$ becomes
\be
X=\frac{1}{3n \ln\tilde{R} }.
\ee
Then, the corresponding expression for $f(R)$ is

\be\label{eq:fclass2}
f(R)=-\frac{6F_0H_0^2\tilde{R}^2}{(3n \ln \tilde{R})^{1/n}}\left[1-\frac{2}{3 n^2 \ln \tilde{R}}+\frac{24-36
   n+39 n^2-b_3}{108 n^4 \ln^2 \tilde{R}}+\mathcal{O}(\ln^{-3}\tilde{R})\right].
\ee
In this case it is not possible to obtain $f(R)$ that are well defined in arbitrary spacetime since $\ln\tilde{R}$ cannot be removed. Therefore, this class can be used only on spacetimes with positive $\tilde{R}$. Further constraints might be necessary if fractional powers of are present.  

A simple example can be obtained by fixing the $b_i$ so the terms in the squared bracket of~\eqref{eq:fclass2} form a geometric series. In that case we have
\be
f(R)=-\frac{F_0R^2}{24H_0^2\left[2+3\ln\left(\frac{-R}{12H_0^2}\right)\right]}.
\ee

The expressions for power spectra and spectral indices read

\be
\mathcal{P}^{\text{II}}_{\text{t}}=\frac{4C_{\text{t}}^2}{\pi^2}\alpha_k^{6/b_1}\left\{ 1+\frac{12\alpha_k}{b_1}+3\alpha_k^2\left[2 \left(1-\gamma _M\right)+\frac{6}{b_1}+\frac{24-b_3}{b_1^2}\right]+\mathcal{O}(\alpha_k^3)\right\},
\ee

\be
n_{\text{t}}^{\text{II}}=-6 \alpha _k^2 \left[1+\alpha\left(2+\frac{b_2}{b_1}\right)\right]+\mathcal{O}(\alpha_k^4),
\ee

 \be
\begin{split}
\mathcal{P}^{\text{II}}_{\text{s}}=\frac{C_{\text{s}}^2\alpha^{6/b_1}}{4\pi^2\alpha^2_k}\left[1-2 \alpha _k \left(1+\frac{b_2-6}{b_1}\right)+\mathcal{O}(\alpha_k^2)\right],
   \end{split}
   \ee
   
\be
n_{\text{s}}^{\text{II}}-1=2\alpha_k^2\left[b_1-3+\alpha _k \left(b_1+2 b_2-6-\frac{3 b_2}{b_1}\right)+\mathcal{O}(\alpha_k^2)\right],
\ee

\be
r^{\text{II}}=48\alpha_k^2\left[1+2 \alpha _k \left(1+\frac{b_2}{b_1}\right) +\mathcal{O}(\alpha_k^2)\right].
\ee

\sect{Change of frame}
\label{sec:frame}
In this section we discuss the change from the Jordan frame to the Einstein frame. In our framework, a change of frame corresponds to a perturbative redefinition of the slow-roll parameter $\alpha$, i.e.
\be
\alpha\rightarrow\tilde{\alpha}(\alpha)=a_1\alpha+a_2\alpha^2+a_3\alpha^3+\ldots .
\ee
In the RG analogy this is equivalent to a change of scheme. Once all the quantities are derived in one frame it is sufficient to apply the redefinition to the appropriate expressions to obtain the results in the new frame. We distinguish the quantities in the Einstein frame by means of a tilde.

The Einstein frame is obtained starting from the action written in the form~\eqref{eq:actphi}

\be
S'(g,\phi)=-\frac{1}{2\kappa^2}\int\sqrt{-g}\left[f(\phi)-F(\phi)\left(\phi-R\right)\right]
\ee
and performing a conformal transformation

\be
g_{\mu\nu}=e^{\hat{\kappa}\chi}\tilde{g}_{\mu\nu}, \qquad \hat{\kappa}\equiv\sqrt{\frac{2}{3}}\kappa,
\ee
where $\chi$ is defined by the constraint

\be
e^{\hat{\kappa}\chi}F(\phi)=1.
\ee
Therefore, we have

\be
 \qquad \mathrm{d}\tilde{s}^2=\tilde{g}_{\mu\nu}\mathrm{d}\tilde{x}^{\mu}\mathrm{d}\tilde{x}^{\nu}=Fg_{\mu\nu}\mathrm{d}x^{\mu}\mathrm{d}x^{\nu}.
\ee
The resulting action in the Einstein frame reads

\be
S_E(\tilde{g},\chi)=\int\sqrt{-\tilde{g}}\left[-\frac{1}{2\kappa^2}R(\tilde{g})+\frac{1}{2}\tilde{g}^{\mu\nu}\partial_{\mu}\chi\partial_{\nu}\chi-U(\chi)\right],
\ee
where

\be
U(\chi)\equiv\frac{e^{2\hat{\kappa}\chi}\left[f(\chi)-\phi(\chi) e^{-\hat{\kappa}\chi}\right]}{2\kappa^2}.
\ee
In order to map the background quantities of the two frames into each other we need a time reparametrization $\tilde{t}(t)$ given by

\be\label{eq:ttilde}
\frac{\mathrm{d}\tilde{t}}{\mathrm{d}t}=\frac{\tilde{a}}{a}=\sqrt{F(\bar{R})},
\ee
where $\bar{R}$ is the background scalar curvature. Form now on every quantity is understood to be computed on the background metric without using a bar. 
From~\eqref{eq:ttilde} we have
\be
\tilde{H}(\alpha)=\frac{1}{\tilde{a}}\frac{\mathrm{d}\tilde{a}}{\mathrm{d}\tilde{t}}=\frac{1}{\sqrt{F}}\left(H+\frac{\dot{F}}{2F}\right)=\frac{H(\alpha)}{\sqrt{F(\alpha)}}\left(1-\alpha\right).
\ee
In order to match the definition of $\tilde{\alpha}$ with the one used in the Einstein frame in~\cite{Anselmi:2021rye} we choose
\be
\tilde{\alpha}\equiv\frac{\hat{\kappa}}{2\tilde{H}}\frac{\mathrm{d}\chi}{\mathrm{d}\tilde{t}}=-\frac{1}{\left(1-\alpha^2\right)}\frac{\dot{F}}{2HF}=\frac{\alpha}{1-\alpha}
\ee
from which we get
\be
\alpha(\tilde{\alpha})=\frac{\tilde{\alpha}}{1+\tilde{\alpha}}.
\ee
Interestingly enough, the redefinition is exact. The beta function in the Einstein frame reads
\be
\beta_{\tilde{\alpha}}(\tilde{\alpha})=(1+\tilde{\alpha})^2\beta_{\alpha}(\alpha(\tilde{\alpha}))=b_n\tilde{\alpha}^{2+n}+\mathcal{O}(\tilde{\alpha}^{3+n}).
\ee
Therefore, functions $f(R)$ of a certain class in the Jordan frame are mapped into potential of the same class in the Einstein frame. Another interesting fact is that in class I the ratio 

\be
\frac{H}{\sqrt{F}}=\frac{H_0}{\sqrt{F_0}}\left[1+\mathcal{O}(\alpha)\right]
\ee
becomes regular in $\alpha=0$. Therefore, in class I also $\tilde{H}(\tilde{\alpha})$ is regular in $\tilde{\alpha}=0$. In fact, we have

\be
\tilde{H}(\tilde{\alpha})=\tilde{H}_0\left[1+\frac{3 \tilde{\alpha }}{b_0}-\frac{3 \tilde{\alpha }^2 \left(b_1-3\right)}{2 b_0^2}+\mathcal{O}(\tilde{\alpha}^3)\right], \qquad \tilde{H}_0\equiv \frac{H_0}{\sqrt{F_0}},
\ee
which matches with the one obtained in~\cite{Anselmi:2021rye}. Moreover, the expression for $\varepsilon$ in the Einstein frame is exact and gives
\be
\tilde{\varepsilon}(\tilde{\alpha})=3\tilde{\alpha}^2
\ee
as expected.

\sect{Class 0: other polynomial $f(R)$}
\label{sec:class0}

We conclude by discussing a special class of theories. The reader might have noticed that the simplest $f(R)$'s we can think of, i.e. polynomials, do not appear in any of the classes studied above, with the exception for the Starobinsky model and pure quadratic gravity in class I. The reason is that simple polynomial extensions of Starobinsky or Hilbert actions have a linear beta function. It is not possible to export the techniques from asymptotically-free quantum field theory and apply our RG analogy at the present stage. In particular, there are no leading logarithms that can be resummed into the running slow-roll parameter. Moreover, as discussed in~\autoref{sect:fRG}, in the assumption $\alpha\ll 1$ a linear beta function is sufficient to guarantee $\varepsilon \ll 1$ but leads to a constant $\eta_H$ [plus corrections of order $\mathcal{O}(\alpha)$] and slow-roll inflation cannot be realized. Nevertheless, this class still admits a systematic expansion in powers of $\alpha$ for the background quantities, since we only require $\alpha\ll 1$ to perturbatively solve the equations. In other words, theories with a linear beta function can still produce a phase of accelerated expansion, but the slow-roll condition $|\eta_H|\ll 1$ is violated, at least far enough in the past (see the discussion below).

In what follows we derive $f(R)$ assuming a linear beta function following the same steps used in~\autoref{sec:classes} to show that any polynomial other than the Starobinsky model and pure quadratic gravity, belongs to this class. Then, we derive the leading-order predictions using standard techniques.

We introduce a linear beta function

\be
\bar{\beta}_{\alpha}=\bar{b}\alpha+\beta_{\alpha}, \qquad \bar{b}<0.
\ee
Then, solving~\eqref{eq:veps} and~\eqref{eq:FHeqs} we find

\be
v=1-\alpha -\frac{3 \alpha ^2}{1+2 \bar{b}}-\frac{3 \alpha ^3 \left(1+\bar{b}-2 b_0\right)}{\left(1+2
   \bar{b}\right) \left(1+3 \bar{b}\right)}+\mathcal{O}(\alpha^4),
\ee

\be
F(\alpha)=F_0\left[1+\frac{2 \alpha }{\bar{b}}+\alpha ^2 \left(\frac{1}{\bar{b}}+\frac{2-b_0}{\bar{b}^2}\right)+\mathcal{O}(\alpha^3)\right],
\ee

\be
H(\alpha)=H_0\left[1+\alpha  \left(1+\frac{1}{\bar{b}}\right)+\alpha ^2 \left(1+\frac{3}{\bar{b}}+\frac{1-b_0}{2
   \bar{b}^2}\right)+\mathcal{O}(\alpha^3)\right].
\ee
In this case de Sitter spacetime is an exact solution because $H(0)=H_0$ and $R(0)=-12H_0^2$. Therefore, the inverse of $R(\alpha)$ is a power series in $\left(1+\frac{R}{12H_0^2}\right)$. We find

\be
\alpha(R)=X \left\{1+\frac{X}{\left(\bar{b}-4\right) \left(\bar{b}+1\right)}\left[14-2 \bar{b}^2+\frac{4-2 b_0}{\bar{b}}-\bar{b}
\left(1+b_0\right)\right]+\mathcal{O}(X^2)\right\},
\ee
where 

\be
X\equiv \frac{2 \bar{b}}{\left(\bar{b}-4\right) \left(\bar{b}+1\right)}\left(1+\frac{R}{12H_0^2}\right).
\ee
Finally, the function $f(R)$ is

\be
f(R)=-6 F_0 H_0^2 \left[1-2 \left(1+\frac{R}{12 H_0^2}\right)-\frac{4\left(1+\frac{R}{12
   H_0^2}\right)^2} {\left(\bar{b}-4\right) \left(\bar{b}+1\right)}+\mathcal{O}\left(\left(1+\frac{R}{12
   H_0^2}\right)^3\right)\right].
\ee
With appropriate choices of the coefficients we can obtain any polynomial extensions of the Starobinsky model, such as
\be
f(R)=R-\frac{1}{6M^2}R^2+\frac{\lambda_3}{M^4}R^3, \qquad f(R)=R-\frac{1}{6M^2}R^2-\frac{\lambda_4}{M^6}R^4
\ee
and so on. The same $f(R)$ without the $R^2$ term can also be obtained, as well as the Starobinsky model with the ``wrong sign" in front of the $R^2$ term plus a positive cosmological constant, i.e.
\be
f(R)=R
+\frac{1}{6M^2}R^2+2\Lambda.
\ee
However, such theory propagates a scalar tachyon and is not physically acceptable.

The result just found shows an important fact about the Starobinsky model. Among all the polynomial $f(R)$ extensions of general relativity, The Starobinsky model is the only one that allows for slow-roll inflation during the initial phase of the evolution of the universe. Moreover, since all the other polynomial $f(R)$ belong to a different class, calculations cannot be obtained by suitably modifying those of the Starobinsky model, not even perturbatively. This can be seen from the fact that, like in the other classes, $F$ and $H$ are not analytic in the first nonvanishing coefficient of the beta function, in this case $\bar{b}$. 

Besides their simple form, polynomials are particularly important because, in general relativity, they are generated by renormalization (together with all other possible scalar quantities built with the Riemann tensor, its covariant derivatives and the metric tensor). This means that they are always present and even if we consider them as small corrections, they cannot be viewed as such, at least in the context of slow-roll inflation. The only way to avoid their radiative generation is to consider renormalizable theories of gravity. The simplest one is obtained by adding the Weyl-tensor-squared term to Starobinsky action. Such theory, called Stelle gravity~\cite{Stelle:1976gc}, violates unitarity due to the presence of ghost degrees of freedom. However, they can be removed by means of a different quantization prescription that turns them into \emph{purely virtual particles}~\cite{Anselmi:2018ibi, Anselmi:2018tmf}. The modifications of Starobinsky predictions due to the presence of purely virtual particle has been derived in~\cite{Anselmi:2020lpp}. See also~\cite{Piva:2023bcf} for a recent review.

As anticipated, $\varepsilon$ is guaranteed to be small, while $\eta_H$ is not. In fact, from equations~\eqref{eq:epseta} we get
\be
\varepsilon=(1+\bar{b})\alpha+\mathcal{O}(\alpha^2), \qquad \eta_H=-\bar{b}+\mathcal{O}(\alpha).
\ee
Therefore, the condition $|\eta_H|\ll 1$ is violated unless $\bar{b}$ is small, which could depend on the free parameters of the theory. However, even if $\bar{b}<1$ there would be a time sufficiently far in the past where $\alpha\ll\bar{b}$, because $\alpha$ tends to zero in the infinite (conformal) past. It seems that an initial phase of the so-called \emph{constant-roll} inflation~\cite{Motohashi:2014ppa,Motohashi:2017vdc} be inevitable. 
We also highlight that even if we choose $\bar{b}$ small enough to mimic slow-roll inflation, the results would be quantitatively different from those of, say, class I where the linear term in the beta is absent. Indeed, the power spectra are nonanalytic in $\bar{b}=0$ and choosing a very small $\bar{b}$ could make higher-order terms larger. However, as we show below, the inverse powers of $\bar{b}$ appear at the NL and NNL order for tensor and scalar power spectra, respectively.

Another potential issue with a linear beta function is the running of $\alpha$. In the case of quadratic and cubic beta functions, the LLO solutions~\eqref{eq:alpharun12} can be expanded in powers of $\alpha_k$, obtaining $\alpha=\alpha_k+\mathcal{O}(\alpha_k^2)$, where the corrections contain powers of $\ln\eta$. This means that in those cases we can approximate $\alpha$ as a constant at the leading order in $\alpha_k$. The same cannot be done in the case of a linear beta function. In fact

\be\label{eq:alpharun0}
\frac{\mathrm{d}\alpha}{\mathrm{d}\ell}=\bar{b}\alpha \qquad\Rightarrow \qquad \alpha(\eta)=\alpha_k\eta^{\bar{b}}.
\ee
If $\bar{b}$ is small enough we could expand~\eqref{eq:alpharun0} and approximate $\alpha$ as a constant. However, if we do so, the inverse powers of $\bar{b}$ that appear at higher orders could jeopardize the perturbative expansion and the reliability of the results. For these reasons this class of theories should be studied with care, in order to avoid mistakes. For example, a more appropriate treatment would be to consider the ratio $\alpha/\bar{b}$ and perform different expansions in the two limiting cases $\alpha/\bar{b}\ll 1$ and $\alpha/\bar{b}\gg 1$. We postpone a detailed study of this class to a future work. For the moment we give the power spectra at the leading order and show the NL orders assuming $\alpha$ constant to illustrate what we explained above.
 
We start by noticing that a linear beta function introduces a constant term in $\sigma_{\text{s}}$. In fact, using~\eqref{eq:sigmat} and~\eqref{eq:sigmas} we get
\be
\sigma_{\text{t}}=\mathcal{O}(\alpha^2), \qquad \sigma_{\text{s}}=\bar{b}(\bar{b}-3)+\mathcal{O}(\alpha).
\ee
We recast the Mukhanov-Sasaki action~\eqref{eq:MSact} in the form
\be\label{eq:MSact0}
S_{\text{t,s}}(w)=\frac{1}{2}\int\mathrm{d}\eta\left[w'^2-w^2+\frac{\nu^2-1/4}{\eta^2}w^2\right],
\ee
where
\be
\nu^2\equiv\sigma+\frac{9}{4}.
\ee
The corresponding equation of motion can be solved for a generic $\nu$. Using the Bunch-Davies initial condition, the solution is given by the well-known Hankel function of the first kind

\be
w(\eta)=e^{\frac{i\pi}{4}(1+2\nu)}\frac{\sqrt{\pi\eta}}{2}H_{\nu}^{(1)}(\eta),
\ee
which in the superhorizon limit is

\be
w(\eta)\simeq -i\frac{e^{\frac{i\pi}{4}(1+2\nu)}}{\sqrt{2\pi}}\left(\frac{\eta}{2}\right)^{\frac{1}{2}-\nu}\Gamma(\nu),\qquad \eta\rightarrow 0,
\ee
where $\Gamma$ is the Euler function. The power spectra are given by

\be
  \mathcal{P}_{\text{t}}=8\frac{k^3}{\pi^2}|u|^2, \qquad \mathcal{P}_{\text{s}}=\frac{k^3}{2\pi^2}|\mathcal{R}|^2,
\ee
where $u$ and $\mathcal{R}$ can be expressed in terms of $w_{\text{t,s}}$ through the relations~\eqref{eq:tensred} and~\eqref{eq:scalred}. Finally, using~\eqref{eq:sigmat}
and~\eqref{eq:sigmas} we have
\be
\nu_{\text{t}}=\frac{3}{2}+\frac{\alpha^2(3-\bar{b})}{1+2\bar{b}}+\mathcal{O}(\alpha^3), \qquad \nu_{\text{s}}=\left|\frac{3}{2}-\bar{b}\right|+\frac{\alpha(\bar{b}-1)(\bar{b}+b_0)}{3-2\bar{b}}+\mathcal{O}(\alpha^2).
\ee
Note that in the scalar case the coefficient $\bar{b}$ appears already at the leading order. 

Putting everything together we find

\be
\mathcal{P}_{\text{t}}=\frac{\kappa^2vH^24^{\nu_{\text{t}}}\Gamma(\nu_{\text{t}})^2}{F\pi^3}\eta^{3-2\nu_{\text{t}}}, \qquad \mathcal{P}_{\text{s}}=\frac{\kappa^2vH^2(1-\alpha^2)4^{\nu_{\text{s}}}\Gamma(\nu_{\text{s}})^2}{48F\pi^3\alpha^2}\eta^{3-2\nu_{\text{s}}},
\ee
which, at $k=k_*$, to the next-to-leading order give

\be
\mathcal{P}_{\text{t}*}=\frac{4C_{\text{t}}^2}{\pi^2}\left[1+\alpha_* ^2 \left(\frac{3}{\bar{b}}-4+\frac{10-8 \gamma _M}{1+2 \bar{b}}+2 \gamma _M\right)+\mathcal{O}(\alpha_*^3)\right],
\ee

\be\label{eq:Psclass0}
\mathcal{P}_{\text{s}*}=\frac{C_s^2 \Gamma \left(\frac{3}{2}-\bar{b}\right)^2}{4^{\bar{b}} \pi ^3\alpha_*^2}\left\{1-2\alpha_*\left[1+\frac{3 \left(1-\bar{b}\right) \left(\bar{b}+b_0\right) }{3-2 \bar{b}}\bigg(\ln 2+\psi
   ^{(0)}\left(3/2-\bar{b}\right)\bigg)\right]+\mathcal{O}(\alpha_*^2)\right\},
\ee

\be
n_{\text{t}*}=-\frac{2 \alpha_* ^2 \left(3-2 \bar{b}\right)}{1+2 \bar{b}}+\mathcal{O}(\alpha_*^2), \qquad n_{\text{s}*}-1=2 \bar{b}-\frac{6 \alpha_*  \left(\bar{b}-1\right) \left(\bar{b}+b_0\right)}{3-2 \bar{b}}+\mathcal{O}(\alpha_*^2),
\ee

where $\psi^{(0)}(x)$ is the digamma function. The tensor-to-scalar ratio at the leading order is
\be
r_*=48\alpha_*^2\left(\frac{2^{\bar{b}-1}\sqrt{\pi}}{\Gamma(3/2-\bar{b})}\right)^2+\mathcal{O}(\alpha_*^2)
\ee
If $\bar{b}$ contains a free parameter of the theory then it always possible to fix it so the experimental constraints are reproduced. However, since $n_{\text{s}*}-1$ is small, this would force $\bar{b}$ to be small as well and the problem of having large corrections might reappear. This should be checked model by model. 

Finally, we derive the number of e-folds. Again, the linearity of the beta function makes the results different from the other classes. In fact, from~\eqref{eq:Nk} we get 
\be
N_*=\frac{\ln\alpha_*}{\bar{b}}+\mathcal{O}(\alpha_*^0).
\ee
To summarize, this class should be studied separately from the others, and they should not be seen as smooth modifications of theories in other classes.

\sect{Conclusions}
\label{sec:conclusions}

We have studied inflationary cosmology in the context of $f(R)$ theories. The various models can be classified by means of a mathematical analogy between the evolution of the background metric during inflation and the renormalization-group flow of quantum field theory. All the results are computed in the Jordan frame and expressed as expansions in a small parameter $\alpha$. A key quantity is the function $\beta_{\alpha}$, which is a power series in $\alpha$ that must be at least quadratic in order to realize slow-roll inflation. The classification is determined by the first nonvanishing term of $\beta_{\alpha}.$ This method helps us in identify the models that cannot explain the cosmic microwave background data. In particular,  we have shown that only two classes of models can satisfy the experimental constraints. Among them, the Starobinsky model stands out as the only $f(R)$ polynomial extension of general relativity that can realize slow-roll inflation. All other polynomial $f(R)$ belong to a special class of models with a linear $\beta_{\alpha}$. The consequence of this property is that they can only realize constant-roll inflation. Since the Starobinsky model and any of its polynomial extension belong to two different classes they cannot be analytically related in their parameter space not even perturbatively. The reason is that two competing expansions should be considered: one in the slow-roll parameter $\alpha$ and one in the coefficient $\bar{b}$ of the linear term in the beta function.

The classification presented here gives general NLL and NNLL formulas for the tensor and scalar power spectra, respectively, that can be used for any $f(R)$ model that belongs to class I or II, once the first few coefficients of $\beta_{\alpha}$ are derived. On a more speculative side, our results point out that principles typically used in particle physics could be relevant for selecting out theories that are suitable for primordial cosmology, given the fact that, once again, the Starobinsky model emerges for its uniqueness.

\appendix
\renewcommand{\thesection}{\Alph{section}} \renewcommand{\theequation}{%
\thesection.\arabic{equation}} \setcounter{section}{0}

\sect{Explicit examples}
\label{app:examples}
Here we collect some explicit formulas for theories of class I. The expansions can be computed at arbitrarily high order. Here we truncate them at different orders for aesthetic reasons\\
1) Starobinsky model\\
\be
f(R)=R-\frac{1}{6M^2}R^2
\ee
is obtained from the beta function

\be
\beta_{\alpha}=-2 \alpha ^2-\frac{5 \alpha ^3}{3}-\frac{65 \alpha ^4}{9}-\frac{1111 \alpha ^5}{27}-\frac{24689 \alpha
   ^6}{81}-\frac{75125 \alpha ^7}{27}+\mathcal{O}(\alpha^8),
\ee
which gives

\be
F=\frac{2}{3\alpha}\left[1-\frac{\alpha }{6}-\frac{\alpha ^2}{9}-\frac{5 \alpha ^3}{27}-\frac{4 \alpha ^4}{27}-\frac{40 \alpha
   ^5}{81}+\frac{16 \alpha ^6}{81}-\frac{8018 \alpha ^7}{2187}+\mathcal{O}(\alpha^8)\right],
\ee

\be
H=\frac{M}{\sqrt{6\alpha}}\left[1-\frac{7 \alpha }{12}-\frac{77 \alpha ^2}{288}-\frac{883 \alpha ^3}{3456}-\frac{50045 \alpha
   ^4}{165888}-\frac{311731 \alpha ^5}{663552}-\frac{7616963 \alpha ^6}{15925248}+\mathcal{O}(\alpha^7)\right],
\ee

\be
\varepsilon=\alpha +\alpha ^2+\frac{4 \alpha ^3}{3}+\frac{16 \alpha ^4}{9}+\frac{77 \alpha ^5}{27}+\frac{313 \alpha
   ^6}{81}+\frac{226 \alpha ^7}{27}+\mathcal{O}(\alpha^8),
\ee

\be
\eta_H=2 \alpha +\frac{5 \alpha ^2}{3}+\frac{23 \alpha ^3}{9}+\frac{70 \alpha ^4}{27}+\frac{530 \alpha
   ^5}{81}+\frac{49 \alpha ^6}{81}+\frac{10145 \alpha ^7}{243}+\mathcal{O}(\alpha^8),
\ee
2)Nonpolynomial function

\be
f(R)=-\frac{R^2}{6 M^2}\left(1+\frac{3}{2}\frac{M^4}{M^4+R^2}\right)
\ee
is obtained from the beta function

\be
\beta_{\alpha}=-4 \alpha ^2+\frac{22 \alpha ^3}{3}-\frac{382 \alpha ^4}{9}+\frac{5486 \alpha ^5}{27}-\frac{290953 \alpha
   ^6}{81}+\frac{640139 \alpha ^7}{54}+\mathcal{O}(\alpha^8),
\ee
which gives

\be
F=\frac{1}{3\sqrt{\alpha}}\left[1-\frac{17 \alpha }{12}+\frac{391 \alpha ^2}{288}-\frac{33401 \alpha ^3}{3456}+\frac{7851307 \alpha
   ^4}{165888}-\frac{116178191 \alpha ^5}{221184}+\mathcal{O}(\alpha^6)\right],
\ee

\be
H=\frac{M}{2\sqrt{3}\alpha^{1/4}}\left[1-\frac{11 \alpha }{24}-\frac{755 \alpha ^2}{1152}-\frac{13157 \alpha ^3}{3072}+\frac{25387315 \alpha
   ^4}{2654208}-\frac{12693236173 \alpha ^5}{63700992}+\mathcal{O}(\alpha^6)\right],
\ee

\be
\varepsilon=\alpha -\alpha ^2+\frac{31 \alpha ^3}{3}-\frac{247 \alpha ^4}{9}+\frac{8213 \alpha ^5}{27}-\frac{192916 \alpha
   ^6}{81}+\frac{1585685 \alpha ^7}{54}+\mathcal{O}(\alpha^8),
\ee

\be
\eta_H=-4 \alpha +\frac{46 \alpha ^2}{3}-\frac{1150 \alpha ^3}{9}+\frac{21644 \alpha ^4}{27}-\frac{655171 \alpha
   ^5}{81}+\frac{14883485 \alpha ^6}{162}+\mathcal{O}(\alpha^7),
\ee
3) Fractional power of $R$

\be
f(R)=R-\frac{1}{6M^2}R^2-\frac{\lambda}{M}(-R)^{3/2},\qquad \lambda>0,
\ee
is obtained from the beta function

\be
\beta_{\alpha}=-\alpha ^2+\alpha ^3 \left(\frac{13}{12}-\frac{2}{\lambda ^2}\right)+\alpha ^4
   \left(\frac{317}{144}+\frac{8}{\lambda ^4}-\frac{47}{3 \lambda ^2}\right)+\mathcal{O}(\alpha^5),
\ee
which gives

\be
F=\frac{\lambda^2}{3\alpha^2}\left[1-\alpha  \left(\frac{25}{6}-\frac{4}{\lambda ^2}\right)+\alpha ^2 \left(\frac{2}{9}-\frac{4}{\lambda
   ^4}+\frac{16}{3 \lambda ^2}\right)+\mathcal{O}(\alpha^3)\right],
\ee

\be
H=\frac{M\lambda}{2\sqrt{3}\alpha}\left[1-\alpha  \left(\frac{49}{12}-\frac{2}{\lambda ^2}\right)-\alpha ^2 \left(\frac{5}{288}+\frac{4}{\lambda
   ^4}-\frac{35}{6 \lambda ^2}\right)+\mathcal{O}(\alpha^3)\right],
\ee

\be
\varepsilon=\alpha +2 \alpha ^2+\alpha ^3 \left(\frac{49}{12}-\frac{2}{\lambda ^2}\right)+\alpha ^4
   \left(\frac{1181}{144}+\frac{8}{\lambda ^4}-\frac{47}{3 \lambda ^2}\right)+\mathcal{O}(\alpha^5),
\ee

\be
\eta_H=\alpha -\alpha ^2 \left(\frac{1}{12}-\frac{2}{\lambda ^2}\right)-\alpha ^3
   \left(\frac{593}{144}+\frac{8}{\lambda ^4}-\frac{41}{3 \lambda ^2}\right)+\mathcal{O}(\alpha^4).
\ee

\bibliographystyle{JHEP} 
\bibliography{mybiblio}

\end{document}